\documentclass[journal, twoside, web, letterpaper]{ieeecolor}
\usepackage{lcsys}
\usepackage[hyphens]{url}
\usepackage{graphicx}

\usepackage{amsthm}
\usepackage{amsmath}
\usepackage{amssymb}
\usepackage{amsfonts}
\usepackage{cite}

\usepackage{amsmath}
\usepackage[dvipsnames]{xcolor}

\usepackage[colorlinks=true, linkcolor=blue]{hyperref}
\usepackage{cleveref}

\newtheorem{theorem}{Theorem}[]

\newtheorem{lemma}{Lemma}[]

\newtheorem{definition}{Definition}[]
\newtheorem{assumption}{Assumption}[]

\crefname{assumption}{assumption}{assumptions}
\Crefname{assumption}{Assumption}{Assumptions}

\newcommand{\Rb}{\mathbb{R}}

\newcommand{\bw}{\mathbf{w}}

\newcommand{\x}{\mathbf{x}}

\newcommand{\A}{\mathbf{A}}
\newcommand{\Ac}{\mathcal{A}}

\newcommand{\cE}{\mathcal{E}}
\newcommand{\Eb}{\mathbb{E}}
\newcommand{\Fc}{\mathcal{F}}

\newcommand{\Pb}{\mathbb{P}}

\newcommand{\I}{\mathbf{I}}
\newcommand{\M}{\mathbf{M}}

\newcommand{\bv}{\mathbf{v}}

\newcommand{\OLS}[2]{\widehat #1^{\textsf{OLS}}_{#2}}

\newcommand{\blend}[2]{\widehat #1^{\textsf{OLS-SME}}_{#2}}

\newcommand{\CLS}[2]{\widehat #1^{\textsf{CLS}}_{#2}}

\usepackage{booktabs} 
\usepackage{multirow}
\usepackage{diagbox}

\title{
System Identification Under Bounded Noise: \\ Optimal Rates Beyond Least Squares
\author{Xiong Zeng, Jing Yu, and Necmiye Ozay 
\thanks{This work was supported by ONR CLEVR-AI MURI (\#N00014-21-1-2431). XZ and NO are with the University of Michigan, Ann Arbor. JY is with the University of Washington, Seattle. Correspondence: {\tt\small zengxion@umich.edu} }
}
}

\begin{document}

\maketitle
\thispagestyle{empty}

\begin{abstract}
System identification is a fundamental problem in control and learning, particularly in high-stakes applications where data efficiency is critical. Classical approaches, such as the ordinary least squares estimator (OLS), achieve an $O(1/\sqrt{T})$ convergence rate under Gaussian noise assumptions, where $T$ is the number of samples. This rate has been shown to match the lower bound. However, in many practical scenarios, noise is known to be bounded, opening the possibility of improving sample complexity. In this work, we establish the minimax lower bound for system identification under bounded noise, proving that the $O(1/T)$ convergence rate is indeed optimal. We further demonstrate that OLS remains limited to an $\Omega(1/\sqrt{T})$ convergence rate, making it fundamentally suboptimal in the presence of bounded noise. Finally, we instantiate two natural variations of OLS that obtain the optimal sample complexity. 
\end{abstract}



\section{Introduction}

\IEEEPARstart{S}{ystem}  identification plays a crucial role in modern control design, especially in applications where accurate models of unknown dynamical systems must be learned from data. In high-stakes and safety-critical systems, where data collection can be costly or risky, sample efficiency is of particular importance. While classical results in system identification provide asymptotic convergence guarantees, they often fail to capture the finite-sample behavior. As a result, recent efforts have focused on analyzing the sample complexity of common system identification methods \cite{simchowitz2018learning,oymak2019non,li2023learning,foster2020learning,sattar2022finite}.  

A fundamental system identification problem is to estimate the unknown system parameter  $\A\in \Rb^{n\times n}$ for an autonomous linear time-invariant (LTI) system:
\begin{equation}
\mathbf{x}_{t+1} =\mathbf{A} \mathbf{x}_t   + \mathbf{w}_t,
\label{lti}
\end{equation}
where $\mathbf{x}_t \in \mathbb{R}^{n}$ and $\mathbf{w}_t \in \mathbb{R}^{n}$ are the state and the noise at time $t$. When the noise $\mathbf{w}_t$ are independent and identically distributed  (i.i.d.) Gaussian random variables, it has been shown that the ordinary least squares estimator (OLS) achieves the optimal convergence rate of $O(1/\sqrt{T})$ (see, e.g., \cite{jedra2020finite}). Consequently, many learning-based control methods have leveraged OLS as a core system identification subroutine, enabling stability, safety, and performance guarantees \cite{dean2020sample, simchowitz2020naive, kargin2022thompson, lale2022reinforcement, zhou2024simultaneous}.

On the other hand, in many applications, system designers have prior knowledge on the noise characteristics. 
Therefore, alternative system identification approaches seek to harness this information to improve sample efficiency. 
Among these, set membership estimation (SME)  algorithms leverage noise boundedness for estimation \cite{bai1998convergence,akccay2004size, kitamura2005size, bai1995membership}. 
One of the key advantages of SME is its ability to provide consistent uncertainty set estimation with convergence guarantees \cite{hespanhol2020statistical}, whereas OLS fails to do so  {for irregular explosive systems}\cite{phillips2013inconsistent,sarkar2019near}. 
Moreover, Li et al. \cite{li2023learning} recently show that {a version of} SME breaks through the $\Omega(1/\sqrt{T})$ convergence rate lower bound attained by OLS for Gaussian noise, achieving a significantly faster $O(1/T)$ convergence rate when the noise has bounded support.

Motivated by \cite{li2023learning}, in this paper, we derive a minimax convergence rate lower bound for system identification when $\mathbf{w}_t$ is i.i.d. zero-mean with bounded support. We prove that indeed {$\Omega(1/T)$} is the minimax lower bound for stable linear dynamical systems with bounded noise (\Cref{thm_minimax_lower_hard_sys_prob}), establishing that the rate achieved by SME is indeed optimal. Furthermore, we demonstrate that the convergence rate lower bound for OLS remains $\Omega(1/\sqrt{T})$ in this setting, revealing an inherent limitation of OLS for system identification problems with bounded noise (\Cref{thm_lower_bound_least_square}). To put our results in perspective, we summarize some of the related lower bound results, including more traditional ones for linear regression with i.i.d. samples, in Table~\ref{tab:compare-lower-bounds}. 

\begin{table}[t]
    \caption{Convergence Rate Lower Bound (LB) Summary.}
    \label{tab:compare-lower-bounds}
    \centering
    \begin{tabular}{c@{\hspace{3pt}}|c@{\hspace{3pt}}|c|c}
        \toprule
        & & \textbf{Minimax LB} &  \textbf{LB for OLS}  \\
        \midrule
        \multirow{2}{*}{\textbf{Regression}} 
            & Gaussian & $\Omega(1/\sqrt{T})$ (\cite{wainwright2019high})   & $\Omega(1/\sqrt{T})$ (\cite{mourtada2022exact}) \\
            & Bounded  & $\Omega(1/T)$ (\cite{yi2024non}) & $\Omega(1/\sqrt{T})$ (\cite{rudelson2008littlewood})  \\
        \midrule
        \multirow{2}{*}{\textbf{LTI Sys Id}} 
            & Gaussian   & $\Omega(1/\sqrt{T})$ (\cite{jedra2019sample}) & $\Omega(1/\sqrt{T})$ (\cite{tu2024learning}) \\
            & Bounded  & $\Omega(1/T)$ (Thm. \ref{thm_minimax_lower_hard_sys_prob})  & $\Omega(1/\sqrt{T})$ (Thm. \ref{thm_lower_bound_least_square}) \\
        \bottomrule
    \end{tabular}
\end{table}


\textbf{Notation} 
We use lower case, lower case boldface, and upper case boldface letters to denote scalars,
vectors, and matrices, respectively.  
For a vector $\mathbf{x}\in \mathbb{R}^n$, $\| \x\|_{\infty}$ denotes its infinity norm. Identity matrices of dimension $n$ are denoted as $\mathbf{I}_n$. 
We use $\operatorname{diag}(\bv)$ for converting a vector $\bv\in \mathbb{R}^n$ into a diagonal matrix in $\mathbb{R}^{n \times n}$. For a matrix $\mathbf{M}\in \mathbb{R}^{m\times n}$, $M(i,j)$ denotes its element in the $i$th row and the $j$th column, and $\|\M\|_2$ and $\rho(\M)$ denote its spectral norm and spectral radius, respectively. 
We use  $\operatorname{exp}(\cdot)$ for the exponential function.
We use $[n]$ as shorthand for index set $\{1,2,\dots,n\}$. We write
$
f(x)={O}(g(x))
$
if and only if there exist constants $N$ and $C$ such that
$
|f(x)| \leq C|g(x)| 
$ for all $x>N$. Similarly, we write
$
f(x)={\Omega}(g(x))
$
if and only if there exist constants $N$ and $C$ such that
$
|f(x)| \geq C|g(x)| 
$ for all $x>N$. {For a set $\cE$, its complement is denoted by $\cE^\mathcal{C}$.}

\vspace{-0.05in}
\section{Preliminaries}
We consider the problem of estimating the system matrix $\A$ of the autonomous system \eqref{lti} from \textit{single trajectory} data.  
In many practical scenarios, performing estimation and data collection tasks on an unstable open-loop system is often impractical and unsafe. Therefore, it is common in system identification literature to assume the open loop is stable. 
 \begin{assumption}[Open-Loop Stable]
\label{assump:stability}
$\rho\left(\A\right)<1$.
\end{assumption}

In this paper, we are particularly interested in understanding the fundamental limit of system identification under i.i.d. \textit{bounded} noise. We formalize the conditions on the noise in the following:
 \begin{assumption}[Bounded Noise]
\label{bounded_noise}
 The noise satisfies $\|\bw_t\|_{\infty} \leq \bar{w}$ for all $t \geq 0$. Further, $\bw_t$ is i.i.d. across coordinates, with zero mean and covariance matrix $\sigma_w^2 \mathbf{I_n}$.
\end{assumption}

\begin{assumption}[Probability Upper Bound of Approaching Boundary]
\label{no_disapppear_boundary_prob}  
    { There exists $ C_{\bar{w}}>0$ such that for all $\epsilon\in[0,\bar{w}]$ and} for all $1 \leq j \leq n$, we have

$$
   \max \left(\mathbb{P}\left(w_t^{(j)} \leq \epsilon- \bar{w}\right), \mathbb{P}\left(w_t^{(j)} \geq \bar{w}-\epsilon\right)\right)  \leq C_{\bar{w}} \epsilon,
$$
where $w_t^{(j)}$ denotes the $j$th entry of vector $\bw_t$.
\end{assumption}
\noindent {Such $C_w$ always exists for any distribution satisfying Assumption \ref{bounded_noise} with a bounded probability density function (pdf). To see this, note that the probabilities in \Cref{no_disapppear_boundary_prob} are the areas under the pdf near $\bar{w}$. Since the pdf is bounded, one can always upper bound the area under pdf with a rectangular function, the height of which is $C_{\bar w}$. For example, uniform and truncated Gaussian distributions trivially satisfy this assumption.}

For simplicity of the analysis, we will also make the following assumption about the initial condition of the system:
\begin{assumption}[Initial Condition]
    \label{assump:initial_condition}
    The system \eqref{lti} starts with the initial condition $\mathbf{x}_t = 0$.
\end{assumption}




\section{Main results}

\subsection{Minimax Sample Complexity Lower Bound}

Our first result proves that the minimax convergence rate lower bound for the system identification of \eqref{lti} under bounded i.i.d. noise is indeed {$\Omega(1/T)$}, where the estimation error decreases at least linearly over the number of samples. 

\begin{theorem}[Minimax Lower Bound]
\label{thm_minimax_lower_hard_sys_prob}
Fix $\delta \in (0,1)$. Let Assumptions \ref{assump:stability}- \ref{assump:initial_condition} hold. Consider the autonomous system \eqref{lti} and a single trajectory $\{\x_t \}^T_{t=1}$ generated from it.
Let $\Fc_T$ denote the $\sigma$-algebra
generated by $\{\x_t \}^T_{t=1}$ and $\hat{\A}_T$ denote the estimated system matrix from any $\Fc_T$-measurable estimator for the system matrix $\A$. Then, for small enough $\epsilon>0$, it holds that 
\begin{equation}
\label{eq:minimax_prob}
\sup_{\hat{\A}_T} \inf_{ \A\in \mathbb{R}^{n\times n}} \mathbb{P}_\A^T \left( \|\hat{\A}_T - \A \|_2 <   \epsilon \right) > 1-\delta ,
\end{equation}
\vspace{5pt}
 only if $T > \frac{1}{4C_{\bar{w}} \bar{w} \epsilon   }\left(1-\frac{2\delta}{n}\right),$
 where $\mathbb{P}_\A^T$ denotes the randomness generated by \eqref{lti} with system parameter $\A$.
\end{theorem}
The proof of this theorem can be found in \Cref{proof_thm_minimax_lower_hard_sys_prob}. 
\Cref{thm_minimax_lower_hard_sys_prob} says that in order to achieve a fixed estimation error $\epsilon$ with high probability, the number of samples must be larger than {$\Omega(1/\epsilon)$}\footnote{Note that the dependence on $\epsilon$ is tight since Li et al. \cite{li2023learning} provide a matching upper bound.}. In other words, the estimation error scales as {$\epsilon = \Omega(1/T)$}. Unlike systems affected by Gaussian noise, \Cref{thm_minimax_lower_hard_sys_prob} shows that imposing a bounded support assumption on the noise fundamentally alters the achievable sample complexity in system identification, revealing a distinct gap between the unbounded and bounded noise regimes. 

This distinction is significant because prior work has established $O(1/\sqrt{T})$ as the optimal convergence rate for systems under Gaussian noise. This result is rooted in the fact that the KL divergence of two $T$-length system trajectories generated by two different system parameters that differ by $\epsilon$ under Gaussian noise
is $O(T\epsilon^2)$ (see e.g. \cite[Section F.2]{simchowitz2018learning}). In contrast, our proof leverages the total variation (TV) distance of the trajectory distributions generated by two different systems under bounded noise. In particular, we show that the TV distance is $O(T\epsilon)$ (\Cref{lem_TV_Divergence_Scalar_System_general} in \Cref{append:lemma1}). This key difference leads to fundamentally different lower bounds in the bounded and unbounded regimes. Furthermore, since the SME algorithm has been shown to attain this rate, \Cref{thm_minimax_lower_hard_sys_prob} establishes that the SME algorithm is indeed optimal in the data trajectory length.



This raises a critical question: Does the optimal estimator for systems with Gaussian noise, such as OLS, remain optimal when the additional bounded support assumption is imposed? In the next section, we demonstrate that the answer is no.

\vspace{-0.05in}

\subsection{Optimality Gap for OLS}
Given single trajectory  data $\{\mathbf{x}_t\}_{t=1}^T$ generated from \eqref{lti}, we study the sample complexity lower bound of OLS for the estimation of the unknown system matrix $\A$:
\begin{equation}\label{eq:ols}
\OLS{\A}{T} = \arg\min_{\A} \sum_{t=1}^{T-1} \left\| \textbf{x}_{t+1} - \A \textbf{x}_t \right\|_2^2. 
\end{equation}


In what follows, we will show that OLS does \textit{not} achieve the optimal rate for systems under bounded noise. For simplicity of analysis, we will focus on scalar systems.
 \begin{theorem}[Lower Bound of OLS]
\label{thm_lower_bound_least_square}  Consider an  autonomous system \eqref{lti} with a scalar system matrix $a\in (-1,1)$ and noise satisfying Assumptions \ref{bounded_noise}-\ref{assump:initial_condition}. Let $\{x_t \}_{t=1}^T$ be a trajectory generated by this system and $\OLS{a}{T}$ be the estimated system parameter via \eqref{eq:ols}. Then we have that for all $a\in (-1, 1)$ and small enough   $\epsilon>0$,
   \begin{align*}  
   \Pb_a^T\left(|\OLS{a}{T}-a|   < \epsilon    \right)   >   1-\delta,
   \end{align*}
  only if  {
  $$  T \! > \!\!\frac{
  \pi \sigma_w^2(1-a^2)^2}{2\epsilon^2(1+\bar{w}^2/\ln2)} \left(1-\delta -  \frac{C_4 }{T^{1/5}}-\frac{C_5 }{\exp\left(  C_6   T \right)}\right)^2,
   $$}
 where $ C_4$, $C_5$, and $C_6$ are universal positive constants.

 \end{theorem} 
 The proof and the details of the constants can be found in \Cref{proof_thm_lower_bound_least_square}. \Cref{thm_lower_bound_least_square} shows that in order for OLS to achieve a fixed estimation error $\epsilon$, the number of samples must be larger {$\Omega(1/{\epsilon^2})$}, making the convergence rate {$\Omega(1/\sqrt{T})$}. {While the classical asymptotic results \cite{lennart1999system} may suggest similar dependence, our proof leverages a quantitative version of the central limit theorem to make this intuition precise for the finite sample setting.}
  
\vspace{-0.1in}
\section{Simulation}
\label{sec:simulation}
\Cref{thm_minimax_lower_hard_sys_prob} establishes that the optimal rate for identifying the system parameter of \eqref{lti} is {$\Omega(1/T)$}. 
Notably, Li et al. \cite{li2023learning} show that SME constructs parameter uncertainty sets whose diameters decrease at this optimal rate, where the uncertainty sets are constructed using the data $\{\textbf{x}_t\}_{t=1}^T$ as:
\begin{equation}
\label{eq:sme}
    \!\!\mathcal{P}_T(\bar{w}):= \left\{ \A\in \mathbb{R}^{n\times n}: \|\mathbf{x}_{t}-\A \mathbf{x}_{t-1}\|_{\infty}\leq \bar{w},\,\,\,   \forall t\in [T] \right\}.
\end{equation}
Therefore, we introduce two natural SME-inspired point estimators that are derived from OLS. We will compare the sample complexity of the standard OLS estimator \eqref{eq:ols} against the two OLS-SME hybrid methods, highlighting their optimal convergence behavior.


\textbf{System setup. }We consider \eqref{lti} with $\A \in \mathbb{R}^{4 \times 4}$ where entries of $\A$ are sampled i.i.d. from uniform distribution bounded by $[-5,\,5]$. Then $\A$ is normalized to have $\rho({\A}) = 0.7$ to comply with \Cref{assump:stability}. 
We use the uniform distribution for the noise with $\bar w = 2$ as the noise bound and sample $\textbf{w}_t$ i.i.d. element-wise.

\textbf{Constants in \Cref{thm_minimax_lower_hard_sys_prob}. } To compute the lower bound, we fix the probability in \eqref{eq:minimax_prob} as   {$\delta = 0.01$}. 
For uniform distribution in $[-2,\,2]$, we have $C_{\bar w} = \frac14$ for \Cref{no_disapppear_boundary_prob}. 

\begin{figure}[t]
    \centering\includegraphics[width=0.7\columnwidth]{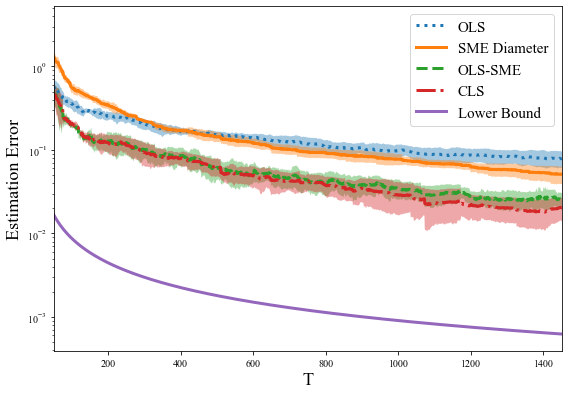}
    \caption{Estimation error convergence for different identification methods}
    \label{fig:result}
\end{figure}

\textbf{System identification methods.} 
We consider two natural SME-based point estimators. The first estimator is named {OLS-SME}, where, after performing OLS, we check whether the generated estimation is inside the SME uncertainty set. If it is outside, we project the OLS estimation on the SME set and call the projected point $\blend{\A}{t}$. Formally,
$$
\blend{\A}{T} := \arg\min_{\A\in\mathcal{P}_T }\|\A - \OLS{\A}{T}\|_2^2.
$$
The second estimator is the constrained least squares estimator, which we denote as {CLS}:
$$
\CLS{\A}{T} := \arg\min_{\A\in\mathcal{P}_T }\sum_{t=1}^{T-1} \left\| \textbf{x}_{t+1} - \A \textbf{x}_t \right\|_2^2.
$$
\textbf{Comparison. } We plot the error\footnote{The code to reproduce the experiment
can be found in \url{https://github.com/jy-cds/Bounded-Noise-SysID-Minimax-Lowerbound.git}.}, which is defined to be the $\ell_2$ distance between the true system parameter and the estimated parameter, for {OLS}, {OLS-SME}, and {CLS} in \Cref{fig:result}. Further, we also plot the diameter of $\mathcal{P}_T$ (``SME diameter''). This represents the worst-case estimation error of any system identification method that constrains the estimated parameter to be inside the SME uncertainty set. 
As predicted by Theorems \ref{thm_minimax_lower_hard_sys_prob} and \ref{thm_lower_bound_least_square}, OLS exhibits a sub-optimal convergence rate while the SME-based methods converge with the same rate as the theoretical lower bound. In particular, { the hybrid methods, like OLS-SME or CLS, can} offer the best of both worlds: they preserve the low estimation error characteristic of OLS { in low-data regime} while simultaneously achieving the optimal convergence rate of SME.

\vspace{-0.06in}
\section{Conclusion}
This work establishes the minimax sample complexity lower bound for system identification under bounded i.i.d. noise, showing that SME-based methods achieve the optimal {$\Omega(1/T)$} convergence rate while the ordinary least squares estimator remains limited to {$\Omega(1/\sqrt{T})$}. Future work includes { improving the dimension and $\delta$ dependence of the lower bound, which is admittedly loose in our current analysis.} It will also be interesting to extend the analysis to more general bounded noise models beyond the infinity norm bound.

\vspace{-0.01in}
\bibliographystyle{IEEEtran}
\bibliography{references}

\appendix
\vspace{-0.01in}
\subsection{TV Distance for Scalar Systems}
\label{append:lemma1}
When the noise is bounded, KL divergence between the distributions over state trajectories of two different systems is, in general, infinity. Therefore, we consider total variation (TV) distance to measure how distinguishable state trajectories are when the system has bounded noise. The TV distance is the largest absolute difference between the probabilities that the two probability measures assign to the same event. 
\begin{definition}[TV Distance] 
\label{def:TVDistance}
Consider a measurable space $(\Omega, \mathcal{F})$, where $\Omega $ is a set and $\mathcal{F}$ is a $\sigma$-algebra on $\Omega $. Consider the probability measures $\mathbb{P}$ and $\mathbb{Q}$ defined on $(\Omega, \mathcal{F})$. Assume  $\mathbb{P}$ and $\mathbb{Q}$ have the pdfs $p(x)$ and $q(x)$ respectively.  The total variation distance between $\mathbb{P}$ and $\mathbb{Q}$ is given by
$$
\operatorname{TV}(\mathbb{P}, \mathbb{Q}) :=\sup _{\cE \in \mathcal{F}}|\mathbb{P}(\cE)-\mathbb{Q}(\cE)|=\frac{1}{2} \int|p(x)-q(x)| d x.
$$
\end{definition}

\begin{lemma}
\label{lem_TV_Divergence_Scalar_System_general}
    Consider two scalar systems $\mathcal{S}_1$ and $\mathcal{S}_2$ of the form \eqref{lti} under Assumptions \ref{assump:stability}-\ref{assump:initial_condition}, where the system parameter is $a_1:= \mu +\epsilon$ and $a_2:= \mu -\epsilon$, respectively, with $ \mu\in (-1+\epsilon,1-\epsilon)$. For $i=1,2$, let $\Pb_{a_i}^T$, $f^T_{a_i}$, and $\mathbb{E}^T_{a_i}$  denote the probability measure of the state trajectory $\{x_t\}^T_{t=1}$, the corresponding probability density function, and the expectation with respect to $\Pb_{a_i}^T$. Then for small enough $\epsilon > 0$, the TV distance between $ \mathbb{P}^T_{a_1}$ and $ \mathbb{P}^T_{a_2}$ satisfies
\begin{equation}
\operatorname{TV}\left(\mathbb{P}^T_{a_1}, \mathbb{P}^T_{a_2}\right) \leq 2C_{\bar{w}}\epsilon \bar{w}  T \frac{1-|\mu|}{(1-|\mu|)^2-\epsilon^2}.
\end{equation}
\end{lemma}
\noindent \textit{Proof: }
    For $i = 1, 2$, let $F_i^{T} :=\prod_{t=1}^{T} f_{a_i}\left(x_{t}  \mid x_{t-1}\right)$.
    \begin{equation*}
\begin{aligned}
    &\operatorname{TV}\left( \Pb^T_{a_1}  ,\Pb^T_{a_2}   \right)  \overset{(a)}{= } \frac{1}{2}\int \dots \int |   f_{a_1}\left(x_T \mid x_{T-1}\right) F_1^{T-1} \\
     & \quad \quad -f_{a_2}\left(x_T \mid x_{T-1}\right) F_2^{T-1}| \mathrm{d} x_T \mathrm{d} x_{T-1}\dots \mathrm{d} x_1\\
     & \overset{(b)}{\leq} \frac{1}{2}\int \dots \int \Big( 2 C_{\bar{w}}\epsilon|x_{T-1}|    F_1^{T-1}  + 2 C_{\bar{w}}\epsilon|x_{T-1}| F_2^{T-1}\\
    & \quad \quad + \left( 1- 2 C_{\bar{w}}\epsilon|x_{T-1}|\right) |    F_1^{T-1}-  F_2^{T-1}|\Big) \mathrm{d} x_{T-1}\dots \mathrm{d} x_1\\
    & =   \Eb_{a_1}^{T-1}[ C_{\bar{w}}\epsilon|x_{T-1}|]+\Eb_{a_2}^{T-1}[ C_{\bar{w}}\epsilon|x_{T-1}| ] \\ 
    &+ {\frac12}\!\int \dots \int        \left( 1-2C_{\bar{w}}\epsilon|x_{T-1}| \right) |    F_1^{T-1}\!-\! F_2^{T-1}|   \mathrm{d} x_{T-1}\dots \mathrm{d} x_1\\
    & \overset{(c)}{\leq}    \Eb_{a_1}^{T-1}[  C_{\bar{w}}\epsilon|x_{T-1}|]+\Eb_{a_2}^{T-1}[  C_{\bar{w}}\epsilon|x_{T-1}| ]  \\
    &\quad \quad + {\frac12}\int \dots \int          |    F_1^{T-1}-  F_2^{T-1}|   \mathrm{d} x_{T-1}\dots \mathrm{d} x_1\\
    &  =    \Eb_{a_1}^{T-1}[  C_{\bar{w}}\epsilon|x_{T-1}|]+\Eb_{a_2}^{T-1}[  C_{\bar{w}}\epsilon|x_{T-1}| ] \\ 
    &\quad \quad +\operatorname{TV}\left( \Pb^{T-1}_{a_1}  ,\Pb^{T-1}_{a_2}   \right) \\
    & \leq  C_{\bar{w}}\epsilon \sum_{t=1}^{T-1} \left( \Eb_{a_1}^{t}[   |x_{t}|]+\Eb_{a_2}^{t}[  |x_{t}| ]\right) \\
    &\overset{(d)}{\leq} 
    C_{\bar{w}}\epsilon \bar{w}  T \left(\frac{1}{1-|\mu+\epsilon|}+ \frac{1}{1-|\mu-\epsilon|}\right)\\
     &\leq 2C_{\bar{w}}\epsilon \bar{w} T \frac{1-|\mu|}{(1-|\mu|)^2-\epsilon^2} ,
    \end{aligned}
\end{equation*}
 where $(a)$ follows from the definition of TV distance and the Markov property of the LTI system with \Cref{bounded_noise}, as well as \Cref{assump:initial_condition}. Inequality 
$(b)$ is due to the following calculation using \Cref{no_disapppear_boundary_prob} and that $f_{a_i}(x_t|x_{t-1}) = f_{w_{t-1}}(x_{t}-a_ix_{t-1})$ for $i \in \{1,2\}$:
 \begin{equation*}
 \begin{aligned}
  &  \int |   f_{a_1}\left(x_t \mid x_{t-1}\right) F_1^{t-1}-f_{a_2}\left(x_t \mid x_{t-1}\right) F_2^{t-1}| \mathrm{d} x_t \\
  &\leq 2 C_{\bar{w}}\epsilon|x_{t-1}|     F_1^{t-1}  + 2 C_{\bar{w}}\epsilon|x_{t-1}|   F_2^{t-1}\\
    &\quad \quad \quad + \left( 1- 2 C_{\bar{w}}\epsilon|x_{t-1}| \right) |    F_1^{t-1}-  F_2^{t-1}|,\end{aligned}
\end{equation*}
for all $t\geq 1$. In $(c)$, we use the fact that choosing $\epsilon < \frac{1-\max\{|a_1|,|a_2|\}}{2C_{\bar{w}}\bar{w}}$ implies that $2C_{\bar{w}}\epsilon|x_{t}| < 1$ holds for all $t>1$. Finally, $(d)$ is because for any $a\in\{a_1,a_2\}$,
{\small \begin{equation}
  \label{upper_bound_x_t}
\begin{aligned}
    |x_{t}| = \left|\sum^{t-1}_{i=0} a^{t-1-i} w_i \right|\leq  \sum^{t-1}_{i=0} |a|^{i} |w_i|<\frac{\bar{w}}{1-|a|}  \, . 
    \end{aligned}
    \end{equation}  }

\vspace{-18pt}
\subsection{Proof of Theorem \ref{thm_minimax_lower_hard_sys_prob}}
\label{proof_thm_minimax_lower_hard_sys_prob}

First, we let $\bv \in \{+1,-1 \}^n$ and define the set of matrices  $\mathcal{A}_\epsilon := \{\A\in \Rb^{n\times n}:\A = \mu \I_n + \epsilon \operatorname{diag}(\bv)\}$ with $\epsilon \in (0,1)$, and $ \mu\in (-1+\epsilon,1-\epsilon)$. For any estimation procedure that outputs $\hat{\A}_T$ as the estimation, define a quantized version $\tilde{\A}_T$ as follows:
\vspace{-8pt}
\begin{equation}
    \tilde{A}_T{(i,j)} = \begin{cases}
        0, \quad i\neq j\\
       \mu + \epsilon, \quad  i=j \; \text{and} \;\hat{A}_T{(i,j)} \geq \mu\\
       \mu - \epsilon, \quad  i=j \; \text{and} \;\hat{A}_T{(i,j)} < \mu
    \end{cases}.
\end{equation}
\vspace{-8pt}
We use $\mathcal{A}_\epsilon$ and $\tilde{\A}_T$ to lower bound the minimax probability:
\begin{equation}
\begin{aligned}
&\inf_{\hat{\A}_T} \sup_{\A \in \Rb^{n\times n} } \mathbb{P}_\A^T \left( \|\hat{\A}_T - \A \|_2 \geq  \epsilon \right)\\
&\geq \inf_{\hat{\A}_T} \sup_{\A\in \Ac_\epsilon} \mathbb{P}_\A^T  \left(\|\hat{\A}_T - \A \|_2 \geq  \epsilon \right)\\
&\geq \inf_{\tilde{\A}_T} \sup_{\A\in \Ac_\epsilon} \mathbb{P}_\A^T  \left(\|\tilde{\A}_T - \A \|_2 \geq 2\epsilon \right).
\end{aligned}
\end{equation}
Next, for all $i\in [n]$, we define the events $\cE_1^i:= \{ \tilde{A}_T{(i,i)} \neq A{(i,i)} \; \text{and} \; \tilde{A}_T{(k,k)} = A{(k,k)} \; \text{for} \; k\in [n] \; \text{and} \; k\neq i  \}$ and $\cE_2^i:= \{\tilde{A}_T{(i,i)} = \mu + \epsilon \; \text{and} \; \tilde{A}_T{(k,k)} = A{(k,k)} \; \text{for} \; k\in [n] \; \text{and} \; k\neq i  \}$. Let $\A_{\epsilon+}^{(i)}$ denote a fixed matrix from the set $\Ac_\epsilon$ with $A_{\epsilon+}^{(i)}(i,i)=\mu+\epsilon$. Let $\A_{\epsilon-}^{(i)}$ denote the matrix that is equal to $\A_{\epsilon+}^{(i)}$ except that on the $i$th diagonal coordinate, $A_{\epsilon-}^{(i)}(i,i)=\mu-\epsilon$. Clearly, $\|\A_{\epsilon+}^{(i)}-\A_{\epsilon-}^{(i)} \|_2=2\epsilon$. Then,
\begin{equation}
\begin{aligned}
   & \inf_{\tilde{\A}_T} \sup_{\A\in \Ac_\epsilon} \mathbb{P}_\A^T \left(\|\tilde{\A}_T - \A \|_2 \geq 2\epsilon \right) \\
   & = \inf_{\tilde{\A}_T} \sup_{\A\in \Ac_\epsilon} \mathbb{P}_\A^T \left(\|\tilde{\A}_T - \A \|_2 = 2\epsilon \right) \\
   & \overset{(a)}{\geq} \inf_{\tilde{\A}_T} \sup_{\A\in \Ac_\epsilon} \sum_{i=1}^n \mathbb{P}_\A^T \left( \cE^i_1  \right)  \\
    & \overset{(b)}{\geq}   \frac{1}{2} \inf_{\tilde{\A}_T}   \sum_{i=1}^n \big[  \mathbb{P}_{\A_{\epsilon+}^{(i)}}^T  \left( \cE^i_1   \right)  +\mathbb{P}_{\A_{\epsilon-}^{(i)}}^T\left( \cE^i_1   \right) \big] \\
     &  \overset{(c)}{=} \frac{1}{2} \inf_{\tilde{\A}_T}   \sum_{i=1}^n \big[ 1- \big( \mathbb{P}_{\A_{\epsilon+}^{(i)}}^T  \left( \cE^i_2  \right)  -\mathbb{P}_{\A_{\epsilon-}^{(i)}}^T  \left( \cE^i_2   \right) \big) \big] \\
    &  {\geq}   \frac{1}{2} \inf_{\tilde{\A}_T}   \sum_{i=1}^n \big[ 1- \big| \mathbb{P}_{\A_{\epsilon+}^{(i)}}^T  \left( \cE^i_2  \right)  -\mathbb{P}_{\A_{\epsilon-}^{(i)}}^T  \left( \cE^i_2   \right) \big| \big] \\
    &  \overset{(d)}{\geq}  \frac{1}{2} \inf_{\tilde{\A}_T}   \sum_{i=1}^n \big[ 1- \operatorname{TV}\big( \mathbb{P}_{\A_{\epsilon+}^{(i)}}^T   ,\mathbb{P}_{\A_{\epsilon-}^{(i)}}^T  \big) \big] \\
    & \overset{(e)}{\geq} \frac{1}{2} n \left( 1-\operatorname{TV}\left( \Pb^T_{a_1}  ,\Pb^T_{a_2}   \right)\right),
    \end{aligned}
\end{equation}
where $(a)$ is because $\{\cE_1^i \}_{i=1}^n$ are $n$ disjoint events and $\cup_{i=1}^{n}\cE_1^i \subseteq \{\|\tilde{\A}_T - \A \|_2 = 2\epsilon \}$. In $(b)$, we use the average of two points $\A^{i}_{\epsilon+}$ and $\A^{i}_{\epsilon-}$ to lower bound the supremum over all $\mathcal{A}_{\epsilon}$, whereas $(c)$ is based on the facts that $\mathbb{P}_{\A_{\epsilon+}^{(i)}}^T  \left( \cE^i_1   \right)=1-\mathbb{P}_{\A_{\epsilon+}^{(i)}}^T  \left( \cE^i_2   \right)$ and $\mathbb{P}_{\A_{\epsilon-}^{(i)}}^T\left( \cE^i_1   \right) = \mathbb{P}_{\A_{\epsilon-}^{(i)}}^T\left( \cE^i_2   \right)$. Inequality $(d)$ is based on Definition \ref{def:TVDistance} for the TV distance, and $(e)$ is from the noise coordinate independence condition in Assumption \ref{bounded_noise} with $a_1= \mu +\epsilon$  and $a_2= \mu -\epsilon$.
 Finally, by Lemma \ref{lem_TV_Divergence_Scalar_System_general} with $\mu=0$, we have
 \vspace{-3pt}
 \begin{equation}
\label{final_ineq_theorem1}
\begin{aligned}
   & \inf_{\hat{\A}_T} \sup_{\A \in \Rb^{n\times n} } \Pb^T_{\A} \left( \|\hat{\A}_T - \A \|_2 \geq  \epsilon \right) \\
    &\geq \frac{1}{2}  n  \left( 1-2C_{\bar{w}}\epsilon  \bar{w}  T \frac{1}{1-\epsilon^2} \right) \overset{(a)}{\geq} \frac{1}{2}  n  \left( 1-4 C_{\bar{w}}\epsilon \bar{w}  T   \right) .
    \end{aligned}
\end{equation}
where $(a)$ is by making $\epsilon$ small enough and in particular $\epsilon^2 < \frac{1}{2}$.  In \eqref{final_ineq_theorem1}, choosing $\delta  $ less than RHS of $(a)$, considering the complement of the event on the first line, and rearranging terms completes the proof.\hfill $\square$

\vspace{-8pt}
\subsection{Proof of Theorem \ref{thm_lower_bound_least_square}}
\label{proof_thm_lower_bound_least_square}
The proof leverages the following quantitative description of the central limit theorem for self-normalized martingales applied to OLS for data from a single trajectory. 

\begin{lemma}[Berry–Esseen for Self-Normalized Martingale for OLS,{\cite[Theorem 3.2]{fan2018berry}}]
\label{lem_berry_ls} Consider the system in \eqref{lti} with a scalar system parameter $a$ and i.i.d noise $w_t$ and data $\{x_t \}_{t=1}^T$ from a \textit{single trajectory}. Suppose that  {$\Eb[w_t]=0$,  $\Eb\left[w_t^2 \right]=\sigma_w^2$ with $\sigma_w>0$, and $\Eb[|w_t|^{4}]<\infty$.} Then
there exists a positive universal constant $C_1$  such that for all $\beta \in \Rb$,
\begin{equation}
 \label{eq:berry-eseen}
\begin{aligned}
& \left|\Pb\left(\left(\OLS{a}{T}-a\right) \sqrt{\sum_{t=1}^T x_t^2} \leq \beta \sigma_w\right)-\Phi(\beta)\right| \\
&     \leq C_1 \Biggl( 
  \left( \frac{1}{(1 - a^2)^2} + \frac{1}{1 - a^4} \frac{\mathbb{E} \left[ |w_t|^4 \right]}{\sigma_w^4} \right) \frac{1}{T} \\
    & \quad  \quad \quad \quad +  \underbrace{\frac{\mathbb{E}\left[ \left(\sum_{t=1}^{T} (x_t^2 - \mathbb{E}[x_t^2]) \right)^2 \right] }{\left( \sum_{t=1}^{T} \mathbb{E}[x_t^2]\right)^2}}_{\mathbb{T}} 
    \Biggr)^{1/5},
\end{aligned}
\end{equation}
where the probability $\Pb$ is with respect to the randomness of $\{w_t \}_{t=0}^{T-1}$ and $\Phi(\beta)$ is the standard Gaussian cumulative distribution function.
\end{lemma}

In what follows, we use the bound \eqref{eq:berry-eseen} to show that the
probability of the estimation error being less than $\epsilon$  {requires the number of samples $T$ to be larger than   $\Omega(1/\epsilon^2 )$.}
We will do so by bounding key quantities in \eqref{eq:berry-eseen}. In particular, we first upper bound the numerator and lower bound the denominator of $\mathbb{T}$. Then, we show that the quantity $\sqrt{\sum_{t=1}^T x_t^2}$ in the LHS of \eqref{eq:berry-eseen} being large is a low probability event.  {Then via a union bound, an upper bound of the probability of the event that   $|\OLS{a}{T} -a|$ is small is obtained.}

\vspace{7pt}
\noindent \textbf{Step 1: Bounding terms in $\mathbb{T}$. }
We will first provide an upper bound to the numerator in the following lemma. 
\begin{lemma}
\label{lem_E_x_E_x_2}
Consider the scalar system \eqref{lti} under Assumption \ref{bounded_noise} with a single state trajectory $\{x_t \}_{t=1}^T$. Then,
\begin{equation*}
    \mathbb{E} \left[\left(\sum_{t=1}^T \left(x_t^2 - \mathbb{E} [x_t^2]\right)\right)^2\right] \leq   
    \frac{\bar{w}^4(1+a^2)}{(1-a)^4(1-a^2)} T.
\end{equation*}
\end{lemma}
\noindent \textit{Proof: }
For \(s > t\), define $\tilde{x}_{s,t}:=\sum_{j=t}^{s-1} a^{s-1-j} w_j$ which is independent of \(x_t\). Then $\Eb[\tilde{x}_{s,t}]=0$ and $x_s^2$ can be written as 
\begin{equation}
\label{eq_x_s_x_s_t}
x_s^2=(a^{s-t} x_t + \tilde{x}_{s,t})^2 = a^{2(s-t)} x_t^2 + 2a^{s-t} x_t \tilde{x}_{s,t} + \tilde{x}_{s,t}^2.
\end{equation}
  Let $\operatorname{Cov}$ and $\operatorname{Var}$ denote the covariance and the variance, respectively.
The covariance between $x_t^2$ and $x_s^2$ is
\begin{equation}
\label{cov_ct2_xs2}
\begin{aligned}
& \operatorname{Cov}(x_t^2, x_s^2)   \overset{(a)}{=} \operatorname{Cov}\Bigl(x_t^2,\; a^{2(s-t)} x_t^2 + 2a^{s-t} x_t\, \tilde{x}_{s,t} + \tilde{x}_{s,t}^2\Bigr)\\
&\overset{(b)}{=} a^{2(s-t)}\,\operatorname{Cov}(x_t^2, x_t^2) + 2a^{s-t}\,\operatorname{Cov}(x_t^2, x_t\,\tilde{x}_{s,t})  \\
&\quad \quad \quad \quad \quad \quad \quad \quad \quad \quad \,\,\quad \quad \quad +\operatorname{Cov}(x_t^2, \tilde{x}_{s,t}^2) \\
&\overset{(c)}{=} a^{2(s-t)}\, \operatorname{Var}(x_t^2),
\end{aligned}
\end{equation}
where $(a)$ is based on \eqref{eq_x_s_x_s_t}, $(b)$ is from the linearity of covariance, and  $(c)$ is because $
\operatorname{Cov}(x_t^2, x_t\,\tilde{x}_{s,t}) = \operatorname{Cov}(x_t^2, \tilde{x}_{s,t}^2)=0.$ Therefore,
\begin{equation*}
\begin{aligned}
\Eb &\left[  \left(\sum_{t=1}^T \left(x_t^2 - \Eb [x_t^2]\right)\right)^2\right] \\
& = \sum_{t=1}^T \operatorname{Var}(x_t^2)+ 2 \sum_{t=1}^{T-1} \sum_{k=1}^{T-t} \operatorname{Cov}(x_t^2, x_{t+k}^2) \\
& \overset{(a)}{=} \sum_{t=1}^T \operatorname{Var}(x_t^2)+ 2 \sum_{t=1}^{T-1} \sum_{k=1}^{T-t} a^{2k}\, \operatorname{Var}(x_t^2) \\
& \overset{(b)}{\leq} \sum_{t=1}^T \frac{\bar{w}^4}{(1-|a|)^4} + 2 \sum_{t=1}^{T-1} \sum_{k=1}^{T-t} a^{2k} \frac{\bar{w}^4}{(1-|a|)^4}  \\
&\le \frac{\bar{w}^4}{(1-|a|)^4} \sum_{t=1}^T \left( 1 + 2\sum^{\infty}_{k = 1} a^{2k} \right)\\
& \le \frac{\bar{w}^4}{(1-|a|)^4}  \left(1 + \frac{2a^2}{1-a^2}\right)T,
\end{aligned}
\end{equation*}
where $(a)$ is from \eqref{cov_ct2_xs2} and  $(b)$ is because $\operatorname{Var}(x_t^2)=\Eb[(x_t^2 - \Eb[x_t^2])^2]\leq \Eb[x_t^4]<\frac{\bar{w}^4}{(1-|a|)^4}$, for which the last inequality is based on  \eqref{upper_bound_x_t}.
\hfill $\square$

Similarly, a lower bound for the denominator of $\mathbb{T}$ can be established as follows:
\begin{equation}
 \label{upper_E_x_t_2}
  \Eb \left[x_t^2 \right] 
=  \sum^{t-1}_{i=0} a^{2(t-1-i)} \Eb\left[w_i^2    \right] \geq  \Eb\left[w_{t-1}^2    \right]=\sigma_w^2.
\end{equation}

\vspace{-3pt}
\noindent \textbf{Step 2: Lower bounding $\sqrt{\sum_{t=1}^T x_t^2}$.}

\begin{lemma}[{\cite[Theorem 2]{jedra2020finite}}] \label{spectral_scalar_covariance}
 Consider the system \eqref{lti} with a scalar system parameter $a$ and i.i.d sub-Gaussian noise $w_t$ and a single trajectory $\{x_t \}_{t=1}^T$.
       Then, for all $ {\gamma}>0$ and for some universal constants $C_2, C_3>0$, we have that  
\begin{align*}
&\Pb_a^T \left( \sqrt{\sum_{t=1}^T x_t^2} >  \frac{\left(1+K^2 \gamma\right)\sqrt{T}}{\sqrt{1-a^2}}   \right)\leq\\
&2 \exp \left(-C_2  \gamma^2  T (1-|a|)^2+C_3 \right),
\end{align*}
where $K$ is an upper bound of the sub-Gaussian norm of  {the noise} $w_t$, e.g., $K \geq \|w_t\|_{\psi_2} $, with $\|w_t\|_{\psi_2} := \inf \left\{ \kappa > 0 : \mathbb{E}\left[ \exp(w_t^2 / \kappa^2)\right] \leq 2 \right\}$. 
\end{lemma}

Note that any bounded random variable $X$ is sub-Gaussian with
$
\|X\|_{\psi_2} \leq \frac{\|X\|_{\infty}}{\sqrt{\ln 2}}
$  {{\cite[Example 2.5.8 (c)]{vershynin2018high}}}. Therefore, when using Lemma \ref{spectral_scalar_covariance} in the latter text, we replace $K$  with $ \frac{ \bar{w}}{ \sqrt{\ln 2}}  $ and we also choose $\gamma=1$ for simplicity.

\noindent \textbf{Step 3: Final bound.}
 {We are now in a position to revisit Lemma \ref{lem_berry_ls}.
Using the fact that $ \Pb(|X|\leq a) =\Pb(X\leq a) -\Pb(X\leq -a) $ for any continuous random variable $X$ and for all $a>0$, and applying \eqref{eq:berry-eseen} twice for $\beta$ and $-\beta$ with $\beta>0$, we obtain}
\begin{align}
&  \Pb_a^T\left(|\OLS{a}{T}-a| \sqrt{\sum_{t=1}^T x_t^2} \leq \sigma_w  \beta \right)   \nonumber\\
& \leq  \Phi\left( \beta \right)-\Phi\left(-\beta \right)+  2C_1\Bigg(\left(\frac{1}{\left(1-a^2\right)^2}+  \frac{\Eb\left[\left|w_t\right|^{4}\right]}{(1-a^{4})\sigma_w^{4}}\right) \frac{1}{T}\nonumber\\
&  \quad \quad \quad \quad \quad\quad\quad\quad\quad  +\frac{\Eb \left[\left|\sum_{t=1}^T \left(x_t^2 - \Eb [x_t^2]\right)\right|^2\right]}{\left(\sum_{t=1}^T \Eb [x_t^2]\right)^2}\Bigg)^{1 /5}\nonumber\\
& \leq   \sqrt{\frac{2}{\pi}} \beta +  \frac{C_4}{T^{1/5}} ,\label{ineq_a_a_hat_sum}
\end{align}
 {
where the last inequality is implied by the combination of   \eqref{upper_E_x_t_2}, Lemma \ref{lem_E_x_E_x_2},  and the fact that $\Phi(\beta)-\Phi(-\beta) < \sqrt{\frac{2}{\pi}} \beta$ for all $\beta>0$, with $$C_4\! \! := \!\!  2C_1 \! \! \left(\!  \frac{1}{\left(\!1-a^2\right)^2}\! +\! \frac{\bar{w}^4}{ (1-a^{4})\sigma_w^4} \! +\!   \frac{\bar{w}^4(1+a^2)}{(1-|a|)^4(1-a^2)\sigma_w^4}\!  \right)^{\frac{1}{5}}.\! \! \!\!$$}
{ Plugging $\beta = \frac{\epsilon(1+K^2 ) \sqrt{T}}{\sigma_w (1-a^2)}$ into \eqref{ineq_a_a_hat_sum} gives
\begin{equation}
\begin{aligned}
  &\Pb_a^T\left(|\OLS{a}{T}-a| \sqrt{\sum_{t=1}^T x_t^2} \leq  \frac{\epsilon(1+K^2 ) \sqrt{T}}{ (1-a^2)} \right)  \\
  &
\leq   \sqrt{\frac{2}{\pi}} \frac{\epsilon(1+K^2 ) \sqrt{T}}{\sigma_w (1-a^2)} +  \frac{C_4}{T^{1/5}}.\label{ineq_a_a_hat_sum_beta}
\end{aligned}
\end{equation}

Now, we are ready to prove the final bound. Define the events $\cE_3:=\left\{|\OLS{a}{T}-a| \sqrt{\sum_{t=1}^T x_t^2} \geq  \frac{\epsilon(1+K^2 ) \sqrt{T}}{ 1-a^2} \right\}$ and $\cE_4:= \left\{\sqrt{\sum_{t=1}^T x_t^2}     {\leq \frac{ (1+K^2 ) \sqrt{T}}{ 1-a^2} }  \right\}$.
It can be observed that
 \begin{equation}
 \label{eq:final}
 \begin{aligned}
 &\Pb_a^T\left(|\OLS{a}{T}-a|   < \epsilon    \right)  \leq  \Pb_a^T\left(  (\cE_3\cap\cE_4)^\mathcal{C} \right) \\
 &\leq   \Pb_a^T\left( \cE_3^\mathcal{C}  \right)  + \Pb_a^T\left(    \cE_4^\mathcal{C} \right)  \\
  &  \overset{(a)}{\leq}  \frac{\epsilon\sqrt{2}(1+\bar{w}^2 /\ln2) \sqrt{T}}{\sqrt{\pi}\sigma_w (1-a^2)}  +  \frac{C_4}{T^{1/5}} +   C_5 \exp \left(-C_6   T  \right), 
   \end{aligned}
   \end{equation}
where $(a)$ follows from the fact that $ \cE_3^\mathcal{C}$ and $\cE_4^\mathcal{C}$ are the events whose probabilities are bounded in \eqref{ineq_a_a_hat_sum_beta} and Lemma \ref{spectral_scalar_covariance}, respectively, and we let $C_5 := 2\exp(C_3) $, and $C_6:=C_2  (1-|a|)^2$. 
Setting the RHS of the third inequality of \eqref{eq:final} to be less than or equal to $1-\delta$, we see that  $$T  \leq \frac{
  \pi \sigma_w^2(1-a^2)^2}{2\epsilon^2(1+\bar{w}^2/\ln2)} \left(1-\delta -  \frac{C_4 }{T^{1/5}}-\frac{C_5 }{\exp\left(  C_6   T \right)}\right)^2,$$
  implies 
$\Pb_a^T\left(|\OLS{a}{T}-a| < \epsilon    \right) \leq 1-\delta $. 

Therefore, 
$\Pb_a^T\left(|\OLS{a}{T}-a| < \epsilon    \right) > 1-\delta,$
only if $$T > \frac{
  \pi \sigma_w^2(1-a^2)^2}{2\epsilon^2(1+\bar{w}^2/\ln2)} \left(1-\delta -  \frac{C_4 }{T^{1/5}}-\frac{C_5 }{\exp\left(  C_6   T \right)}\right)^2.$$}



\end{document}